# Electronic structures of hexagonal $R$MnO$_3$ ($R$ = Gd, Tb, Dy, and Ho) thin films

W. S. Choi,[1] D. G. Kim,[2] S. S. A. Seo,[1] S. J. Moon,[1] D. Lee,[1] J. H. Lee,[1] H. S. Lee,[2] D.-Y. Cho,[2] Y. S. Lee,[3] P. Murugavel,[1] Jaejun Yu,[2] and T. W. Noh[1,*]

[1]*ReCOE & FPRD, Department of Physics and Astronomy, Seoul National University, Seoul 151-747, Korea*
[2]*CSCMR, Department of Physics and Astronomy, Seoul National University, Seoul 151-747, Korea*
[3]*Department of Physics, Soongsil University, Seoul 156-743, Korea*

We investigated the electronic structure of multiferroic hexagonal $R$MnO$_3$ ($R$ = Gd, Tb, Dy, and Ho) thin films using both optical spectroscopy and first-principles calculations. One of the difficulties in explaining the electronic structures of hexagonal $R$MnO$_3$ is that they exist in nature with limited rare earth ions (i.e., $R$ = Sc, Y, Ho – Lu), so a systematic study in terms of the different $R$ ions has been lacking. Recently, our group succeeded in fabricating hexagonal $R$MnO$_3$ ($R$ = Gd, Tb, and Dy) using the epitaxial stabilization technique [Adv. Mater. **18**, 3125 (2006)]. Using artificially stabilized hexagonal $R$MnO$_3$, we extended the optical spectroscopic studies on the hexagonal multiferroic manganite system. We observed two optical transitions located near 1.7 eV and 2.3 eV, in addition to the predominant absorption above 5 eV. With the help of first-principles calculations, we attribute the low-lying optical absorption peaks to inter-site transitions from the oxygen states hybridized strongly with different Mn orbital symmetries to the Mn $3d_{3z^2-r^2}$ state. As the ionic radius of the rare earth ion increased, the lowest peak showed a systematic increase in its peak position. We explained this systematic change in terms of a flattening of the MnO$_5$ triangular bipyramid.

## I. INTRODUCTION

Multiferroic oxides have attracted considerable recent attention due to their intriguing coupling between the magnetic and electric order parameters.[1-10] This magnetoelectric coupling in a single material could lead to new applications, and new understanding of the underlying physics involved. Of all the known multiferroic materials, the $R$MnO$_3$ rare-earth manganites attracted particular interest. These intriguing material systems have two kinds of crystal structure. Depending on the rare earth ionic radius, they form either an orthorhombic phase ($R$ = Bi, La – Dy) or a hexagonal phase ($R$ = Sc, Y, Ho – Lu). Due to the recent observations of very strong magnetoelectric coupling, there has been a flurry of investigations of orthorhombic $R$MnO$_3$.[5,6,11-14] However, orthorhombic $R$MnO$_3$ has relatively low ferroelectric and magnetic ordering temperatures that are typically well below the temperature of liquid nitrogen.[5,6,11]

On the other hand, hexagonal $R$MnO$_3$ (hexa-$R$MnO$_3$) have ferroelectric properties with fairly large remnant polarization and quite high Curie temperature ($T_C$), typically above 590 K. They also exhibit antiferromagnetic behaviors, but their Neel temperature ($T_N$) is quite low, in the range 70 K-120 K, probably due to geometrical frustration.[7] Their magnetoelectric couplings were reported to be smaller than those of orthorhombic $R$MnO$_3$.[7,8] However, such couplings have been clearly identified in the ferroelectric and antiferromagnetic phases.[7-9,15-17] For example, the magnetic ordering in hexa-$R$MnO$_3$ can be controlled by a static electric field[7] or the static dielectric constant shows anomalies when an external magnetic field is applied.[15,16]

To understand the physics involved in these multiferroic materials, it is necessary to understand their electronic structure. Compared to orthorhombic $R$MnO$_3$,[18,19] the detailed electronic structure of hexa-$R$MnO$_3$ is not very well understood. There have been attempts using first-principle calculations,[20,21] optical spectroscopy,[10,22] x-ray absorption spectroscopy (XAS),[23] photoemission spectroscopy (PES),[24] and second harmonic generation studies.[25] Despite these efforts, controversy remains over the origins of the energy bands of hexa-$R$MnO$_3$ near the Fermi surface.[21,22,24,25]

One of the difficulties in understanding the electronic structure of hexa-$R$MnO$_3$ is the absence of any systematic link between the physical properties and tuning parameters. For example, in the bulk hexa-$R$MnO$_3$ phase, the $T_C$ and $T_N$ values do not show a systematic variation with the radius of the $R$ ion.[26,27] As well, all of the optical conductivity spectra $\sigma(\omega)$ of the bulk hexa-$R$MnO$_3$ show interband optical transitions located nearly at the same frequencies.[10,22] In this sense, it is quite worthwhile to extend the phase space of hexa-$R$MnO$_3$ by synthesizing the metastable hexagonal phases of manganites to search for possible variations in electronic structure. Recently, our group successfully fabricated thin films of hexa-$R$MnO$_3$ ($R$ = Gd,[28] Tb,[9,29] and Dy[30]) using the epitaxial stabilization (epi-stabilization) technique.[31] This film fabrication technique opened the way to fabricating new multiferroic materials as well as extending the phase diagram of hexa-$R$MnO$_3$ phases. We also found that significant variations in physical properties exist in newly-synthesized hexa-$R$MnO$_3$ thin films.

In this paper, we report on the systematic changes in the electronic structure of the hexa-$R$MnO$_3$ ($R$ = Gd, Tb, Dy, and Ho) thin films with variation of the rare-earth ions, using optical spectroscopy and first-principles calculations. Optical spectroscopy has been a powerful tool for investigating the systematic electronic band structure of solids.[32,33] We found that $\sigma(\omega)$ of the hexa-$R$MnO$_3$ films is qualitatively similar to that of the bulk hexa-$R$MnO$_3$ (i.e., $R$ = Lu and Y), including the interband transitions. However, the lowest peak shows a systematic trend, i.e., an increase in peak energy with the increase in $R$ ionic radius. Combined with the results of first-principles calculations, we attribute this systematic change to the change of crystal field energy in the Mn ion by linked to the ionic radius of the $R$ ion. This shows that local crystal distortion could play an important role in changing the electronic structure of the hexagonal phases.

## II. EXPERIMENTAL AND THEORETICAL METHODS

### A. Film fabrication

We used pulsed laser deposition techniques to fabricate high-quality hexa-$R$MnO$_3$ ($R$ = Dy, Tb, Gd, and Ho), and half-substituted (Dy, Ho)MnO$_3$ and (Tb, Ho)MnO$_3$ thin films. Note that in bulk form, DyMnO$_3$, TbMnO$_3$, and GdMnO$_3$ have orthorhombic phases. We used the epi-stabilization technique to convert these materials into hexagonal form. We grew them in the hexagonal phases by depositing the films on single crystalline yttria-stabilized zirconia (YSZ) (111) substrates. [Note that for transmission studies, we used YSZ (111) single crystals polished on both sides as substrates.] Since the atomic arrangements of the substrate surfaces form hexagonal nets, the metastable hexa-$R$MnO$_3$ phase could be formed by maintaining the coherent film-substrate interface and minimizing the surface energy.[31] Using x-ray diffraction (XRD) measurements, we confirmed that all of the thin films were grown epitaxially with their $c$-axis perpendicular to the film surface. More details about the growth condition and structural characterization of the thin films are published elsewhere.[9,28-30]

### B. Optical measurements

We obtained near-normal-incident reflectance and transmittance spectra of the thin films in the photon energy range of 0.1 - 6.0 eV. We used a FT-IR spectrometer (Bruker IFS66v/S) and a grating-type spectrophotometer (CARY 5G) at 0.1 - 1.2 eV and 0.4 - 6.0 eV, respectively. Figure 1 shows the optical spectra of a hexa-TbMnO$_3$ thin film (solid lines) and a YSZ (111) bare substrate (dashed lines). The thick blue and thin red lines represent the transmittance and reflectance spectra, respectively. Since the band-gap edge of the YSZ (111) substrate is located near 5 eV, we were able to obtain meaningful transmittance spectra up to that level. In the spectral range of 0.15 – 5.0 eV, we could determine the in-plane $\sigma(\omega)$ of the hexa-$R$MnO$_3$ thin films from the transmittance and reflectance spectra using a numerical iteration process called the intensity transfer matrix method.[34]

### C. First-principles calculations

We carried out first-principles calculations for the hexa-$R$MnO$_3$ ($R$ = Y and Gd) system. We first performed the band calculations for YMnO$_3$, since it does not have any $f$-electrons, and since its structural parameters have been sufficiently studied for the theoretical calculations.[35] On the other hand, there have been few structural studies of hexa-GdMnO$_3$. To include the atomic positions of GdMnO$_3$ in the calculation, we used experimental neutron diffraction data for YMnO$_3$[35] but with putting the Gd ions in the place of the Y ions. The calculation unit cell contains six formula units of hexa-$R$MnO$_3$. To deal with the effects of strong Coulomb interactions among 3$d$- and 4$f$- electrons, we used the LDA+$U$ methods based on density functional theory (DFT), as implemented in a linear-combination-of-localized-pseudo-atomic-orbital (LCPAO) code.[36-40] We used the effective on-site Coulomb energy parameter $U_{\text{eff,Mn}}$ = 4 eV for Mn 3$d$, which turned out to be suitable for describing the band-gap and magnetic properties of MnO.[36] For the Gd 4$f$-electrons, we used $U_{\text{eff,Gd}}$ = 6 eV. To describe the ground state spin-ordering properly, we performed non-collinear spin calculations. We used Troullier-Martins-type norm-conserving pseduopotentials to replace the deep core potentials. For numerical integration and the solution of Poisson's equation, we used an energy cutoff of 240 Ry. For the $k$-space integrations, we used a 4 x 4 x 4 grid.

We found that the first-principles calculation results of YMnO$_3$ are sufficient for understanding detailed electronic structures of hexa-$R$MnO$_3$ near the Fermi level. For both YMnO$_3$ and GdMnO$_3$, the electronic structures related to the Mn 3$d$ and O 2$p$ manifolds showed nearly identical features near the Fermi level. Since the positions of $f$-electron energy levels are located at about 8 eV below and 10 eV above the Fermi level, the effects of $f$-electrons on the valence electronic structure could be considered as minimal.

### III. RESULTS AND DISCUSSION
#### A. Generic spectral features of hexagonal $R$MnO$_3$

Figure 2 shows the in-plane $\sigma(\omega)$ of hexa-$R$MnO$_3$ ($R$ = Gd, Tb, Dy, and Ho) thin films. Note that the hexa-GdMnO$_3$, TbMnO$_3$, and DyMnO$_3$ films were artificially fabricated using the epi-stabilization technique. For the sake of comparison, $\sigma(\omega)$ of single crystal YMnO$_3$ is also shown.[22] The overall spectral shapes of the thin films and the bulk sample are similar. [Note that similar optical spectra were also reported in the literature for the other bulk hexa-$R$MnO$_3$.[10,22]] This similarity suggests that our epi-stabilized hexa-$R$MnO$_3$ thin films should have a similar electronic structure to that of the bulk hexa-$R$MnO$_3$.

Up to this point, $\sigma(\omega)$ of bulk $R$MnO$_3$ ($R$ = Sc, Lu, Er, and Y) has been interpreted in terms of two absorption peaks. However, we noted that TbMnO$_3$ has three important spectral features, which are indicated by arrows in Fig. 2. Compared with $\sigma(\omega)$ of YMnO$_3$, the second peak near 2.3 eV is much clearer in our thin film samples. This is partly because our transmittance measurements are much more sensitive than the reflectance measurements used for the bulk samples. Thus, we can state that most hexa-$R$MnO$_3$ have three important spectral absorption peak features: a sharp peak near 1.7 eV, a weak peak near 2.3 eV, and the most prominent peak located higher than 5 eV.

Another notable point is that the position of the first peak shows a systematic change between 1.61 eV and 1.81 eV, depending on the $R$ ion. The vertical dotted line in Fig. 2 indicates the first peak position of YMnO$_3$. As the radius of the $R$ ion increases from Y to Gd, the position of the first peak shifts to higher energy. The shift in peak position becomes significant for artificially-fabricated hexa-$R$MnO$_3$ films, especially hexa-GdMnO$_3$. The systematic change of the sharp optical transition peak in hexa-$R$MnO$_3$ implies that there should be systematic variations in their electronic structures depending on the radius of the $R$ ion.

### B. Attribution of the optical absorption peaks

There has been debate on the origin of the sharp optical transition at ~ 1.7 eV. One interpretation is that it comes from the charge transfer transition from the O 2$p$ to the Mn 3$d$ states.[21,22,24] An earlier PES study suggested that while the unoccupied level has Mn 3$d$ character, the highest occupied level has mainly O 2$p$ character.[24] Other Mn 3$d$ states in the valence band are located at much lower energy, so that they cannot contribute to the optical transition at ~ 1.7 eV. The other interpretation is that it comes from the on-site $d$-$d$ transition between the Mn 3$d$ levels.[10,25] Hexa-$R$MnO$_3$ has a Mn-O triangular bipyramid cage, so two $e_{1g}$ orbitals ($d_{yz}$ and $d_{zx}$) form the lowest level, followed by two $e_{2g}$ orbitals ($d_{xy}$ and $d_{x^2-y^2}$), and the $a_{1g}$ orbital ($d_{3z^2-r^2}$), as shown in Fig. 5(a). In this simple atomic picture, the transition at ~1.7 eV could be attributed to the on-site $d$-$d$ transition between the occupied $e_{2g}$ orbitals and the unoccupied $a_{1g}$ orbital.

Our LDA+$U$ calculations for hexa-$R$MnO$_3$ electronic structures provide new insights into this issue. We found that the calculated results for GdMnO$_3$ were quite similar to those for YMnO$_3$, so we will explain the calculation results in detail only for YMnO$_3$. Figure 3 shows the orbital-resolved density of state (DOS) of YMnO$_3$. The unoccupied manganese $d_{3z^2-r^2}$ state is located just ~ 1 eV above the Fermi level. On the other hand, for the occupied states near -2 eV, the in-plane O 2$p$ and all Mn $d$ orbitals except for $d_{3z^2-r^2}$ orbital contribute together. These states can be interpreted in terms of two hybridized states between in-plane O 2$p$ orbitals with the associated Mn $d$ orbitals. The dashed blue line shows Mn DOS with the Mn $d_{yz}$ and $d_{zx}$ orbitals. This state is hybridized with the in-plane O 2$p$ orbitals forming a state with $d_{yz}/d_{zx}$ symmetry. On the other hand, the solid green line shows Mn DOS with the Mn $d_{xy}$ and $d_{x^2-y^2}$ orbitals. This state is also hybridized with the in-plane O 2$p$ orbitals forming a state with $d_{xy}/d_{x^2-y^2}$ symmetry. Note that the Mn DOS of the state with $d_{yz}/d_{zx}$ symmetry is located at somewhat lower energy than that with $d_{xy}/d_{x^2-y^2}$ symmetry.

This calculated electronic structure explains the three spectral features in Fig. 2 quite well. Taking into account the energy levels in Fig. 3, the sharp first absorption peak at ~ 1.7 eV can be attributed to the inter-site optical transition from the occupied hybridized state with the $d_{xy}/d_{x^2-y^2}$ orbitals to the unoccupied Mn $d_{3z^2-r^2}$ state. Similarly, the weak second peak at ~ 2.3 eV can be attributed to the inter-site optical transition from the occupied hybridized state with the $d_{yz}/d_{zx}$ orbitals to the unoccupied Mn $d_{3z^2-r^2}$ state. The strong third peak above 5 eV is the sum of two interband transitions: one from the broad O 2$p$ states at ~ -4 eV to the Mn $d_{3z^2-r^2}$ state, and the other from the O 2$p$ states at ~ -2 eV to the Mn 3$d$ states with various orbitals at ~ 3 eV. In these assumptions, it is worth pointing out that the minute structure near 2.3 eV, which was hardly discernible in the previous studies of the bulk samples, can be properly attributed. The agreement between the experimental spectroscopy data and the theoretical explanations guarantees that first-principles calculation results do represent the electronic structure of the hexa-$R$MnO$_3$ quite reliably.

Our investigations revealed that hybridization could play a crucial role in the electronic structure of hexa-$R$MnO$_3$ materials. To date, the optical transitions that involve an empty $d$-electronic state have been attributed to either $d$-$d$ or $p$-$d$ transitions. Such simple interpretations are not suitable for explaining the electronic structure of

hexa-$R$MnO$_3$ materials. As shown in Fig. 3, the occupied states that are responsible for the peaks at ~ 1.7 eV and ~ 2.3 eV come from the strong hybridization between the O 2$p$ and the related Mn 3$d$ orbitals. Due to the strong hybridized nature of the occupied states, we cannot attribute the optical transition in the hexagonal phase simply to strict $d$-$d$ or $p$-$d$ transitions. Instead, the optical transitions in hexa-$R$MnO$_3$ should be regarded as an inter-site charge transfer excitation from the oxygen states strongly hybridized with associated Mn orbital symmetries to the Mn 3$d_{3z^2-r^2}$ state.

**C. Systematic shift of the inter-site optical transition peak at ~ 1.7 eV**

We now discuss the systematic shift of the optical transition peak at ~ 1.7 eV depending on the $R$ ion in hexa-$R$MnO$_3$. Figure 4(a) shows the peak position as a function of the radius of the $R$ ion. For clarity, we have also included the peak positions for the bulk hexa-$R$MnO$_3$ samples from the literatures[10,22] and the peak positions of artificially fabricated half-substituted (Dy, Ho)MnO$_3$ and (Tb, Ho)MnO$_3$ thin films. Except for the small deviations for LuMnO$_3$ and HoMnO$_3$, an increase in the peak position according to the increase in radius of the $R$ ion is clearly visible. Note that the peak position of the hexagonal manganite with the largest $R$ ion among this series, i.e., GdMnO$_3$, is very high, reaching 1.81 eV. This value is higher by ~ 0.2 eV than those of usual bulk hexa-$R$MnO$_3$, which typically have peak positions around 1.6 eV.

We looked for physical properties that could be related to the observed changes in peak position. Figure 4(b) shows the systematic variation in the lattice constant ratio $a/c$ and the in-plane lattice constant $a$ for hexa-$R$MnO$_3$. The lattice constant values of the bulk samples are taken from the literature,[41] whereas the values of the thin film samples were obtained from our XRD measurements. These two structural quantities increase as a function of the ionic radius of the $R$ ion, similar to the $R$ ionic radius dependence of the ~ 1.7 eV peak position. [Although not shown here, we also found that the $c$-axis lattice constant remains nearly the same, unrelated to the $R$ ionic radius.] This indicates that the change in the electronic structure of hexa-$R$MnO$_3$ should be closely related to the structural changes.

**D. Possible origin of the ~ 1.7 eV peak shift**

To explain the close relationship between the change in the first peak position and variations in the structural properties, we looked at how structural changes might affect the energy levels of hexa-$R$MnO$_3$ in an atomic picture. Although such a simple picture cannot take into account all of the strong hybridization effects, it can still provide some insight into the relationship between the structure and energy bands. Figure 5(a) shows the configuration of the Mn-O triangular bipyramid and associated energy level diagram for Mn 3$d$ electrons. If we assume that the $c$-axis lattice constant remains fixed as we increase the radius of the $R$ ion, the MnO$_5$ triangular bipyramid will flatten out. Then, due to the crystal field splitting, the energy level of the Mn $d_{3z^2-r^2}$ state should increase significantly, while the Mn $d_{xy}/d_{x^2-y^2}$ state would change only very little. Therefore, the optical transition from the Mn $d_{xy}/d_{x^2-y^2}$ state to the Mn $d_{3z^2-r^2}$ state will shift to higher energy as the radius of the $R$ ion increases. This simple atomic picture can explain the systematic increase of the first optical peak with increase of $R$ ionic radius.

As we showed in Section III.B, the energy bands just below the Fermi surface should have very strong hybridizations between O $2p$ and the related Mn $3d$ orbitals. To take into account these hybridization effects properly, we relied on our first-principles calculations. To address the relationship between the electronic structure and the MnO$_5$ triangular bipyramid flattening, we calculated the DOS for a hypothetical material system that has the same crystal structure as that of YMnO$_3$ but with a 2% reduction in spacing between the Mn and the apical oxygen ions.[42] Figure 5(b) shows the DOS of the strongly hybridized Mn $d_{3z^2-r^2}$ state for normal YMnO$_3$ and YMnO$_3$ with the MnO$_5$ triangular bipyramid flattened by 2%. There are few changes in the associated DOS below the Fermi level, but the unoccupied Mn $d_{3z^2-r^2}$ energy level moves by 0.24 eV to higher energy. [We found that there are few changes in the DOS of Mn $3d$ orbitals with other symmetries.] These theoretical results explain the observed increase of the first peak from about 1.6 eV to 1.8 eV. Therefore, the observed spectral position changes in Fig. 4(a) should come from the flattening of the MnO$_5$ triangular bipyramid in the hexa-$R$MnO$_3$ structure.

## IV. SUMMARY

We investigated the optical conductivity spectra of artificially fabricated hexagonal $R$MnO$_3$ ($R$ = Gd, Tb, Dy, and Ho) thin films using optical spectroscopy and first-principles calculations. We were able to attribute the optical transition properly by comparing the experimental and theoretical results. We found that the Mn $3d$ states just below the Fermi energy had strong hybridized characteristics between the O $2p$ and the associated Mn $3d$ orbitals. We also found a systematic increase in the lowest optical transition at ~ 1.7 eV with an increase in the $R$ ionic radius. From these studies, we found that the electronic structure of the hexagonal $R$MnO$_3$ should have a close relationship to the crystal structure, especially the distance between Mn and apical O ions in the MnO$_5$ triangular bipyramid. Investigations of other physical properties of the artificially fabricated hexagonal $R$MnO$_3$ could enhance our understanding of the intriguing multiferroic manganite system.

**ACKNOWLEDGEMENTS**

We acknowledge valuable discussions with J. S. Kang. This study was financially supported by Creative Research Initiatives (Functionally Integrated Oxide Heterostructures) of the Ministry of Science and Technology (MOST) and the Korean Science and Engineering Foundation (KOSEF). YSL was supported by the Soongsil University Research Fund.

**FIGURE CAPTIONS**

FIG. 1. (Color online) Optical spectra of YSZ (111) substrate (dashed lines) and hexagonal TbMnO$_3$ thin film (solid lines) artificially fabricated on YSZ (111) substrate using the epitaxial stabilization technique. Reflectance and transmittance spectra are shown as the thin red and thick blue lines, respectively.

FIG. 2. Optical conductivity spectra of hexagonal HoMnO$_3$, DyMnO$_3$, TbMnO$_3$, and GdMnO$_3$ thin films. For the sake of comparison, the reported spectrum of bulk YMnO$_3$ (Ref. 22) is also shown. For clarity, the spectra have been plotted with offsets of 1250 $\Omega^{-1}$cm$^{-1}$ vertically between each curve, and the base line for each curve is shown by the horizontal dashed line. The arrows in the TbMnO$_3$ spectrum indicate the three peak positions where optical absorption occurs. The vertical dotted line indicates the first optical transition peak position of YMnO$_3$.

FIG. 3. (Color online) Orbital-resolved densities of states of Mn 3$d$ orbitals and the in-plane O 2$p$ orbital for YMnO$_3$.

FIG. 4. (a) Peak positions of the optical transition peak at ~1.7 eV for numerous hexagonal $R$MnO$_3$. Empty squares denote the peak positions of the bulk hexagonal $R$MnO$_3$ ($R$ = Lu (Ref. 10), Er, and Y (Ref. 22)), whereas filled squares denote those of the thin film hexagonal $R$MnO$_3$ ($R$ = Dy, Tb, and Gd). Filled circles denote the peak positions of the half-substituted hexagonal $R$MnO$_3$ thin films ($R$ = (Dy$_{0.5}$Ho$_{0.5}$) and (Tb$_{0.5}$Ho$_{0.5}$)). (b) The lattice constant ratio ($a/c$, filled symbols) and the in-plane lattice constant ($a$, empty symbols) of bulk hexagonal $R$MnO$_3$ (circles, $R$=Lu, Yb, Tm, Er, Y, and Ho (Ref. 41)), and film hexagonal $R$MnO$_3$ (squares, $R$ = Dy, Tb, and Gd). The thick gray lines in both (a) and (b) are included simply as visual aids.

FIG. 5. (Color online) (a) Schematic representation of the crystal field splitting changes due to flattening of the MnO$_5$ triangular bipyramid. The flattening occurs due to the increase of the rare earth ionic radius with the fixed $c$-axis lattice constant. As the flattening occurs, the $d_{3z^2-r^2}$ orbital and the $d_{yz}$ and $d_{zx}$ orbitals shift to higher energy, whereas the $d_{xy}$ and $d_{x^2-y^2}$ orbitals stay the same. (b) First-principles calculation results for the Mn $d_{3z^2-r^2}$ orbital DOS of YMnO$_3$ without and with flattening of the triangular bipyramid. The calculations took into account the hybridization between the in-plane O 2$p$ and associated Mn 3$d$ orbitals. When the spacing between the Mn and the apical oxygen ions is reduced by 2%, the DOS for the unoccupied Mn $d_{3z^2-r^2}$ orbital state shows an upward shift of 0.24 eV.

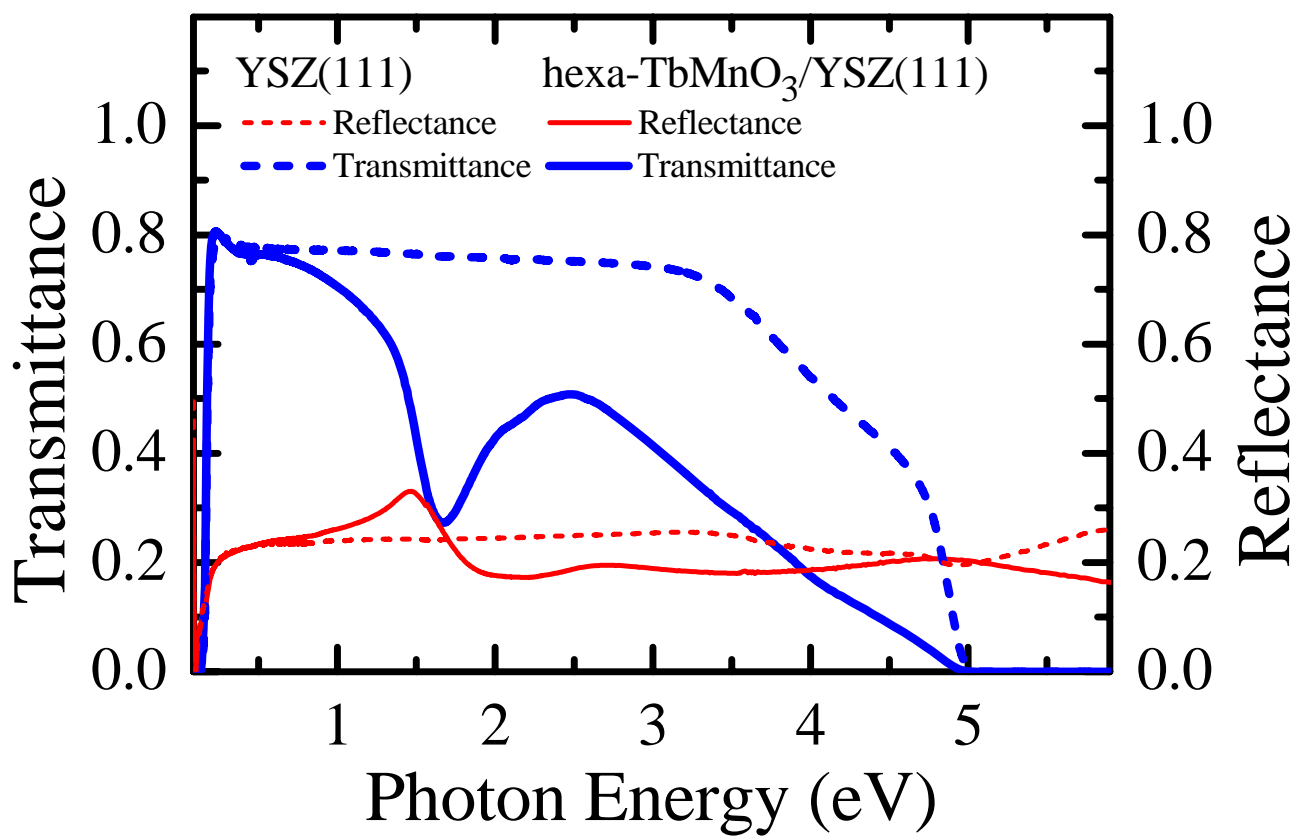

Fig. 1

Choi *et al.*

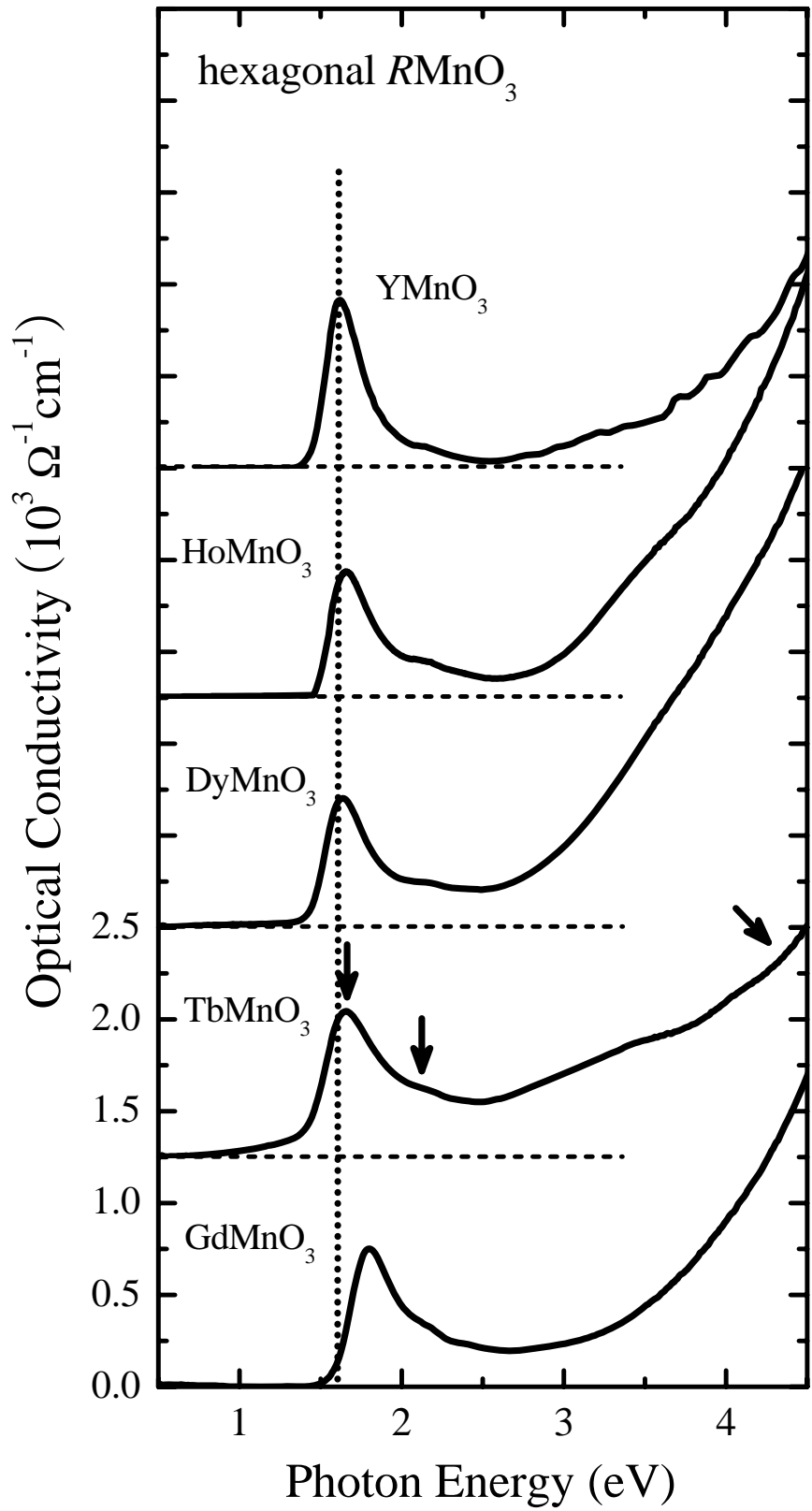

Fig. 2

Choi *et al.*

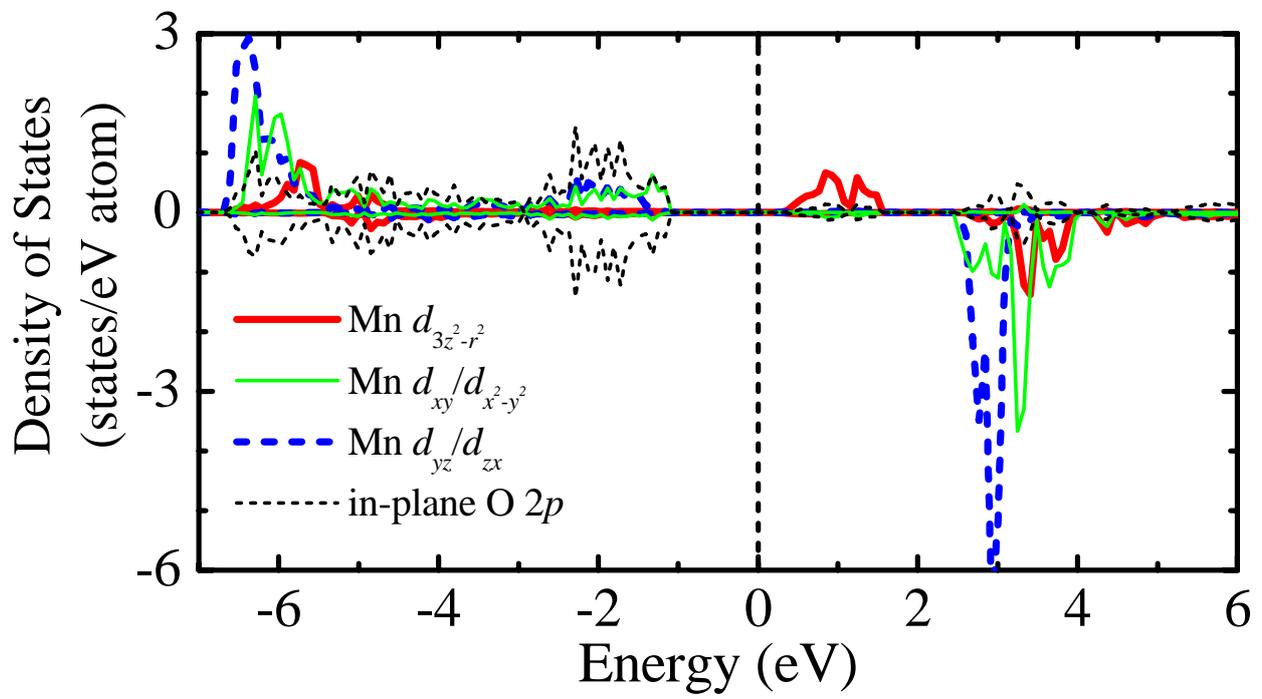

Fig. 3

Choi *et al.*

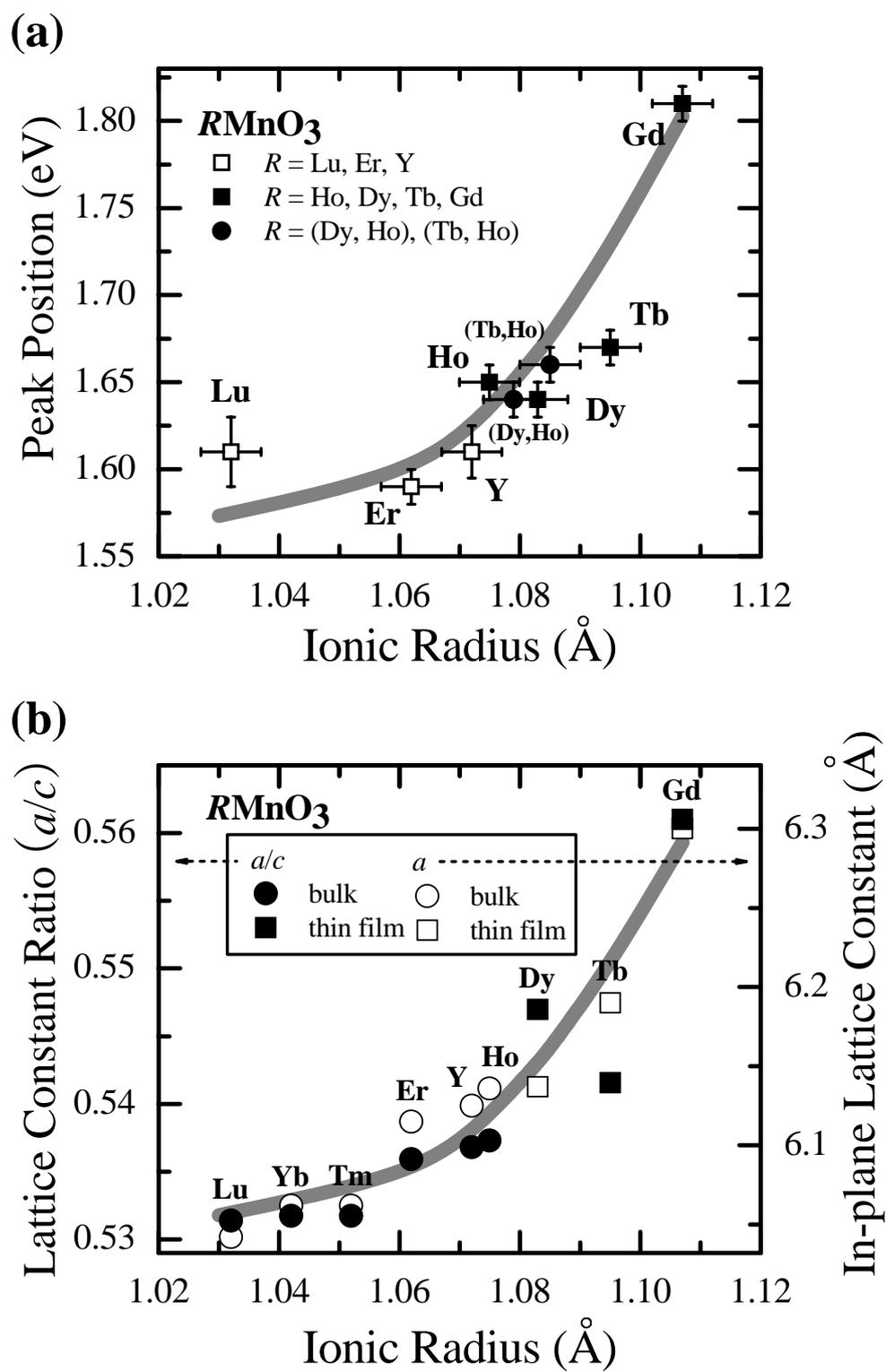

Fig. 4

Choi *et al.*

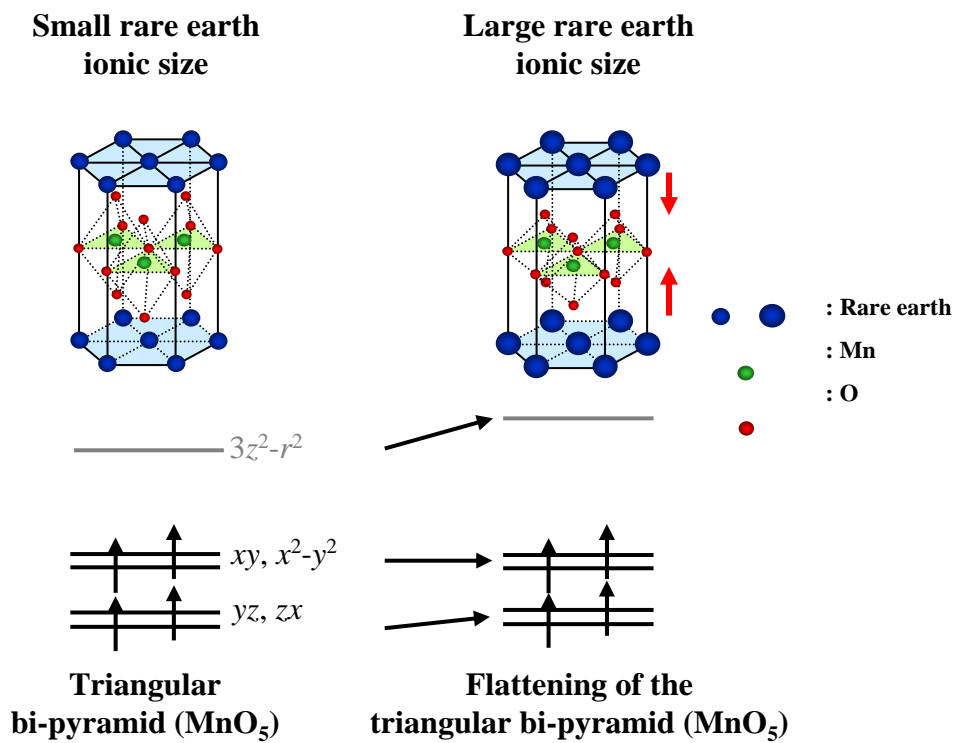

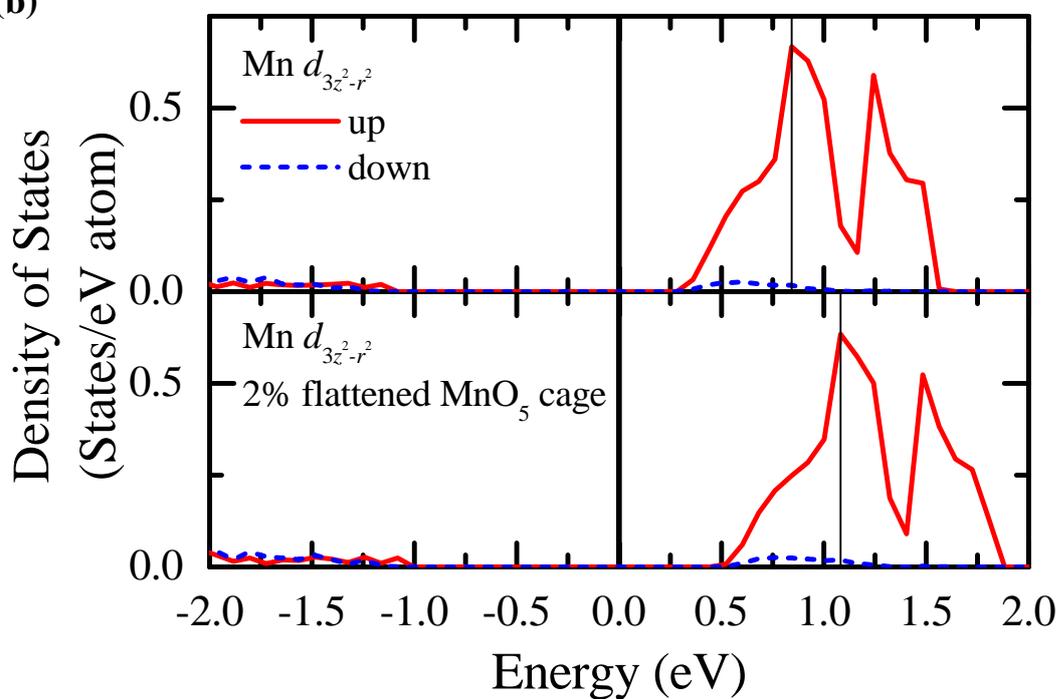

Fig. 5

Choi *et al.*